\documentclass[prc,aps,twocolumn,showpacs,showkeys]{revtex4}
\usepackage{graphicx}

\begin{document}

\newcommand{\p}{\partial} 
\newcommand{\ls}{\left(} 
\newcommand{\rs}{\right)} 
\newcommand{\beq}{\begin{equation}} 
\newcommand{\eeq}{\end{equation}} 
\newcommand{\beqa}{\begin{eqnarray}} 
\newcommand{\eeqa}{\end{eqnarray}} 
\newcommand{\bdm}{\begin{displaymath}} 
\newcommand{\edm}{\end{displaymath}} 

\title{Application of density dependent parametrization models to 
asymmetric
 nuclear matter}


\author{B. Liu$^{1,2}$, M. Di Toro$^{3}$, V. Greco$^{3}$, C.W. Shen$^{1,4}$,
E.G. Zhao$^{5}$, B.X. Sun$^{6}$}
\affiliation{$^{1}$ Center of Theoretical Nuclear Physics, National Laboratory
 of Heavy Ion Accelerator, Lanzhon 730000, China\\
 $^{2}$ Institute of High Energy Physics, Chinese Academy of Sciences,
 Beijing 100049, China \\
 $^{3}$ Laboratori Nazionali del Sud, Via S. Sofia 62,
  I-95123 Catania and University of Catania, Italy \\
$^{4}$ School of Science, Huzhou Teachers College, Huzhou 313000, China\\
$^{5}$ Institute of Theoretical Physics, Chinese Academy of
Sciences, Beijing 100080, China\\
$^{6}$ Institute of Theoretical Physics, College of Applied Sciences, Beijing
 University of Technology,
Beijing 100022, China}

\begin{abstract}

Density dependent parametrization models of the nucleon-meson effective
couplings, including
the isovector scalar $\delta$-field, are applied to asymmetric nuclear matter.
The nuclear equation of state and the neutron star properties
are studied in an effective Lagrangian density approach, using the
relativistic mean field hadron theory.
It is known that the introduction of a $\delta$-meson in the constant coupling
scheme leads to an increase of the symmetry energy at high density and so to
larger neutron star masses, in a pure nucleon-lepton scheme.
We use here a more microscopic density dependent model of the nucleon-meson 
couplings to study the
properties of neutron star matter and to re-examine the $\delta$-field effects
in asymmetric nuclear matter.
Our calculations show that, due to the increase of the effective $\delta$
coupling at high density, with density dependent couplings the neutron star 
masses in fact can be even reduced. 
\end{abstract}

\pacs{21.30.Fe, 21.65.+f, 05.70.C; 97.60.Jd;}

\keywords{Density dependent relativistic hadron field theory;
 Nuclear matter; Equation of state; Neutron stars; Beta Equilibrium.}

\maketitle

\section*{1. Introduction}

The understanding of the properties of nuclear matter, at both normal
and high density regions, is of crucial importance in explaining
the formation and structure of neutron stars after the supernova explosion.
The experiments with unstable nuclear beams and relativistic heavy ions
are potential tools in determining which are the best equations of state
($EOS$) that are able to describe hot and dense matter. The properties of 
neutron stars ($NS$)
are characterized by masses and radii, which are obtained from
an appropriate $EOS$ at high densities. The $EOS$ can be derived either from
relativistic or potential models.

The nonlinear Walecka model ($NLWM$) \cite{SW85,bog97}
and derivative scalar couplings \cite{ZM90}, based
on the relativistic mean-field ($RMF$) approach, have been extensively
used to study the properties of nuclear and neutron matter, $\beta$-stable
nuclei and then extended to the drip-line regions.
In the last years some  authors 
\cite{liubo02,menpro04,baranPR,gait04,liubo05} have stressed the 
importance of including
the isovector scalar virtual $\delta (a_{0}(980))$ field in hadronic effective
field theories for asymmetric nuclear matter. The role of the $\delta$ meson
in isospin channels appears relevant  at high densities
\cite{liubo02,menpro04,baranPR,gait04,liubo05} 
and so of great interest in nuclear astrophysics.

In order to describe the medium dependence of nuclear interactions,
a density dependent relativistic hadron field ($DDRH$) theory
has been recently suggested \cite{fuchs95,TW99,hof01}. The density dependent 
meson-nucleon couplings are based on microscopic Dirac-Brueckner ($DB$)
calculations \cite{jong98,hof01,vandal04}   and
adjusted to reproduce some nuclear matter and finite nuclei properties
\cite{fuchs95,TW99,hof01}.
The main intention of this work is to apply different parametrizations of
the density dependent meson-nucleon couplings, including the $\delta$ meson,
to asymmetric nuclear matter. In particular we will see the predictions of
 the density
dependent coupling models when applied to the neutron stars ($NS$).
In fact it is known that the introduction of $\delta$-meson in
the constant coupling model \cite{liubo05} leads to heavier neutron stars
 in a nucleon-lepton picture. This is not obvious for density dependent
models.

The paper is arranged as follows. In Sect.2 the model formalism
is shortly derived. The meson-nucleon coupling parametrizations
are presented in Sect.3. Results and discussions are given in Sects.4, 5.

\vspace{-1.0cm}
\section*{2. The model formalism}

The  Lagrangian density, with $\delta$ mesons, used in this work reads

\begin{widetext}
\begin{eqnarray}\label{eq:1}
{\cal L } &=& \bar{\psi}[i\gamma_{\mu}\partial^{\mu}-(M-
g_{\sigma}\sigma -g_{\delta}\vec{\tau}\cdot\vec{\delta})
-g{_\omega}\gamma_\mu\omega^{\mu}-g_\rho\gamma^{\mu}\vec\tau\cdot
\vec{b}_{\mu}]\psi \nonumber \\&&
+\frac{1}{2}(\partial_{\mu}\sigma\partial^{\mu}\sigma-m_{\sigma}^2\sigma^2)
+\frac{1}{2}m^2_{\omega}\omega_{\mu} \omega^{\mu}
+\frac{1}{2}m^2_{\rho}\vec{b}_{\mu}\cdot\vec{b}^{\mu} \nonumber
\\&&
+\frac{1}{2}(\partial_{\mu}\vec{\delta}\cdot\partial^{\mu}\vec{\delta}
-m_{\delta}^2\vec{\delta^2}) -\frac{1}{4}F_{\mu\nu}F^{\mu\nu}
-\frac{1}{4}\vec{G}_{\mu\nu}\vec{G}^{\mu\nu},
\end{eqnarray}
\end{widetext}

\noindent
with the isoscalar (scalar,vector) $\sigma,\omega_\mu$ and isovector 
 (scalar,vector) $\delta,\rho_\mu$, named $\vec{\delta},\vec{b}_{\mu}$,
 effective fields.
$F_{\mu\nu}\equiv\partial_{\mu}\omega_{\nu}-\partial_{\nu}\omega_{\mu}$
 and  $\vec{G}_{\mu\nu}\equiv\partial_{\mu}\vec{b}_{\nu}-
\partial_{\nu}\vec{b}_{\mu}$.

The most important difference to conventional $RMF$ theory is the contribution
from the rearrangement self-energies to the $DDRH$ baryon field equation.
The meson-nucleon couplings $g_{\sigma}$,
$g_{\omega}$, $g_{\rho}$ and $g_{\delta}$ are assumed to be vertex functions
of Lorentz-scalar bilinear forms of the nucleon field operators.
In most applications of $DDRH$ theory, these couplings are chosen as functions 
of the
vector density $\hat{\rho}^{2}=\hat{j}_{\mu}\hat{j}^{\mu}$ with
$\hat{j}_{\mu}=\bar{\psi} \gamma_{\mu}\psi$.

The equation of state ($EOS$) for nuclear matter at T=0 is 
obtained from the energy-momentum tensor. In a mean field approximation
the energy density has the form \cite{TW99,hof01}

\begin{eqnarray}\label{eq:13}
&&\epsilon=\sum_{i=n,p}{2}\int \frac{{\rm d}^3k}{(2\pi)^3}E_{i}^\star(k)
+\frac{1}{2}m_\sigma^2\sigma^2
+\frac{1}{2} \frac{g_{\omega}^2}{m_\omega^2}\rho^2
\nonumber \\
&&+\frac{1}{2}\frac{g_{\rho}^2}{m_{\rho}^2}\rho_3^2
+ \frac{1}{2}\frac{g_{\delta}^2}{m_{\delta}^2}\rho_{s3}^2,
\end{eqnarray}

\noindent
and the pressure is

\begin{eqnarray}\label{eq:14}
&&p =\sum_{i=n,p} \frac{2}{3}\int \frac{{\rm d}^3k}{(2\pi)^3}
\frac{k^2}{E_{i}^\star(k)}
 -\frac{1}{2}m_\sigma^2\sigma^2
 +\frac{1}{2}\frac{g_{\omega}^2}{m_\omega^2}\rho^2
\nonumber \\
&&+\frac{1}{2}\frac{g_{\rho}^2}{m_{\rho}^2}\rho_3^2
-\frac{1}{2}\frac{g_{\delta}^2}{m_{\delta}^2}\rho_{s3}^2-\Sigma_{o}^{R}\rho,
\end{eqnarray}%
\noindent
with ${E_i}^\star=\sqrt{k^2+{{M_i}^\star}^2},~i=p,n$. The nucleon
effective masses are
${M_p}^\star=M-g_\sigma\sigma-g_\delta\delta_3$ and 
${M_n}^\star=M-g_\sigma\sigma+g_\delta\delta_3$, where the scalar fields,
$\sigma$ (isoscalar) and $\delta_3$ (isovector) are expressed in terms of the
corresponding local scalar densities.  In the pressure a 
rearrangement term appears, in the density dependent cases, as
\begin{eqnarray}\label{sigmazero}
&&\Sigma_0^{R}=(\frac{\partial g_\sigma}{\partial \rho})\frac{g_\sigma}
{m_\sigma^2}\rho_{s}^2
+(\frac{\partial g_\delta}{\partial \rho}) \frac{g_\delta}
{m_\delta^2}\rho_{s3}^2 \nonumber \\
&&-(\frac{\partial g_\omega}{\partial \rho}) 
\frac{g_{\omega}}{m_{\omega}^2} \rho^2
-(\frac{\partial g_\rho}{\partial \rho}) 
\frac{g_{\rho}}{m_{\rho}^2} \rho_{3}^{2},
\end{eqnarray}
\noindent
where
$\rho_3=\rho_p-\rho_n$ and $\rho_{s3}=\rho_{sp}-\rho_{sn}$,  
$\rho_i$ (i=p,n) and $\rho_s$ are
the nucleon and the scalar densities, respectively.


The chemical potentials for protons and neutrons can be written as, 
respectively

\begin{equation}\label{eq:16}
\mu_p=\sqrt{k_{F_i}^2+{{M_p}^\star}^2}
+ \frac{g_{\omega}^2}{m_{\omega}^2}\rho
+\frac{g_{\rho}^2}{m_{\rho}^2}\rho_{3}-\Sigma_{o}^{R},
\end{equation}

and

\begin{equation}\label{eq:17}
\mu_n=\sqrt{k_{F_i}^2+{{M_n}^\star}^2}
+ \frac{g_{\omega}^2}{m_{\omega}^2}\rho
-\frac{g_{\rho}^2}{m_{\rho}^2}\rho_{3}-\Sigma_{o}^{R},
\end{equation}
\noindent
where the Fermi momentum $k_{F_i}$ of the nucleon is related to its
density, $k_{F_i}=(3\pi^2 \rho_i)^{1/3}$.

Since we are interested in the effects of the Nuclear $EOS$
we will consider only pure nucleonic (+leptons) neutron
star structure, i.e. without strangeness bearing baryons and even
deconfined quarks, see \cite{LP04,maieron04}.
The composition is determined by the requirements of charge neutrality and
$\beta$-equilibrium.

The chemical equilibrium condition for a $(n,p,e^{-})$ system
can be written as


\begin{equation}\label{eq:20}
\mu_{e} =\mu_{n}-\mu_{p}= 4E_{sym}(\rho)(1-2X_{p})~,
\end{equation}

\noindent
where $X_{p}$ is the proton fraction $\rho_p/\rho$. 
The symmetry energy can be obtained from the energy per nucleon 
in the parabolic approximation :

\begin{equation}\label{eq:24}
E_{sym}(\rho)=[E/A(\rho,\alpha)-E/A(\rho, \alpha=0)]/\alpha^2,
\end{equation}
\noindent
where $\alpha$ asymmetry perameter $\alpha \equiv (N-Z)/A = -\rho_3/\rho$.
Since the electron density $\rho_{e}$ in the ultra-relativistic
limit for non-interacting electrons can expressed as a function of
its chemical potential,
the charge neutrality condition is

\begin{equation}\label{eq:21}
\rho_{e} = \frac{1}{3\pi^2} \mu_{e}^{3}=\rho_{p}=X_{p}\rho~.
\end{equation}

\noindent
Then, for a given $\rho$,  the $X_{p}$ is related to the nuclear 
symmetry energy by

\begin{equation}\label{eq:23}
3\pi^{2}\rho X_{p} - [4E_{sym}(\rho)(1-2 X_{p})]^{3}=0.
\end{equation}


In the case of the $(n,p,e^{-},\mu^{-})$ system, the constituents of
neutron stars are neutrons, protons, electrons and muons. The threshold
 density for the appearance of muons is when the electron chemical
potential is larger than the muon rest mass : $\mu_{e}>m_{\mu}$=106.55 MeV.
The chemical  equilibrium for the $(n,p,e^{-},\mu^{-})$ system reads

\begin{equation}\label{muemu}
\mu_{\mu}=\mu_{e}= \mu_{n}-\mu_{p} .
\end{equation}

\noindent
The charge neutrality condition is

\begin{equation}\label{rhopemu}
\rho_{p} =\rho_{e}+\rho_{\mu},
\end{equation}

\noindent
with the muon density $\rho_{\mu}$ expressed as a function of its chemical 
potential

\noindent
\begin{eqnarray}\label{rhomu}
\rho_\mu=\frac{1}{3\pi^2}(\mu_\mu^2-m_\mu^2)^{3/2}\theta(\mu_e-m_\mu).
\end{eqnarray}

The proton fraction $X_{p}$ for ($npe$) and ($npe\mu$) systems can be obtained 
by  solving Eq.(\ref{eq:23}) and 
Eqs.(\ref{muemu},\ref{rhopemu},\ref{rhomu}), respectively.
The $EOS$ for the $\beta$-stable ($npe$) and ($npe\mu$) matter can be estimated
by using the obtained values of $X_{p}$.
The equilibrium properties of the neutron stars can be finally studied by
solving Tolmann-Oppenheimer-Volkov ($TOV$) equations  \cite{T39,OV39}
inserting the derived nuclear $EOS$ as an input.
We note that the presence of muons slightly increases the proton fraction 
for a fixed density, making the matter softer. We will see this effect 
in the final equilibrium properties.


\par
\vspace{0.3cm} \noindent

\begin{center}
{{\large \bf Table 1.}~~Parameters of the model }.

\par
\vspace{0.5cm} \noindent

\begin{tabular}{c|c|c|c|c|c} \hline
      &\multicolumn{2}{c}{$TW~[10]$} &$DDRH\rho$ 
&\multicolumn{2}{c}{$DDRH\rho\delta$} \\ \hline
 Meson          &$\sigma$  &$\omega$   &$\rho$    &$\rho$    &$\delta$ 
\\ \hline
$m_i~(MeV)$     &550       &783        &770       &770       &980     
 \\ \hline
$g_i(\rho_{0})$ &10.73  &13.29   &3.59     &5.86    &7.59     \\ \hline
$a_i$           &1.36  &1.40   &0.095  &0.095  &0.02   \\ \hline
$b_i$           &0.23  &0.17   &2.17     &2.17     &3.47    \\ \hline
$c_i$           &0.41  &0.34   &0.05   &0.05   &-0.09   \\ \hline
$d_i$           &0.90  &0.98   &17.84   &17.84   &-9.81    \\ \hline
\multicolumn{6}{c}{$\rho_0=0.153~ fm^{-3}$}            
                 \\ \hline
\end{tabular}
\end{center}



\begin{figure}[hbtp]
\begin{center}
\includegraphics[scale=0.38]{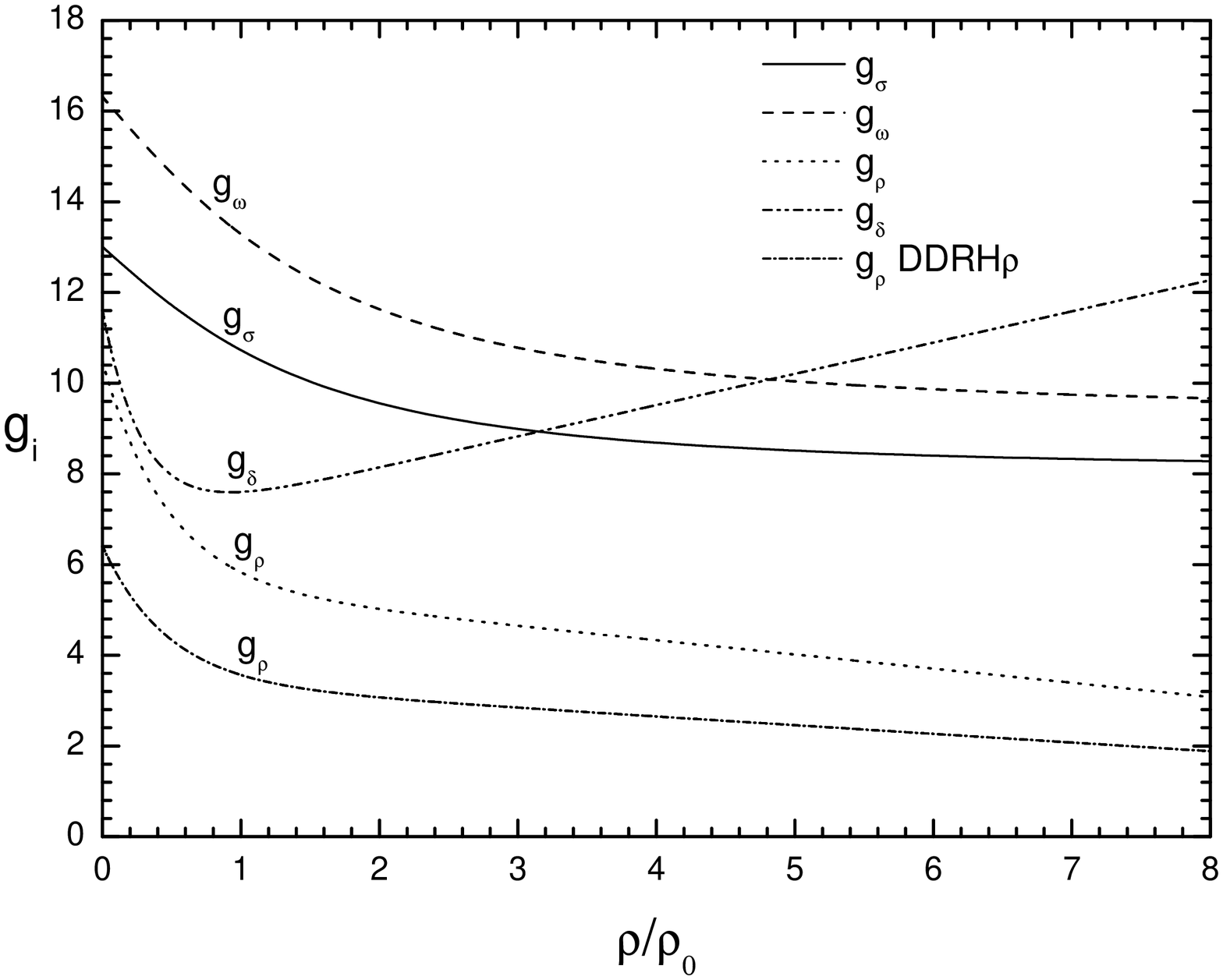}
\vglue -0.5cm
\caption{Density dependence of the meson-nucleon couplings.}
\label{gcoup}
\end{center}
\end{figure}

\vspace{-1.0cm}
\section*{3. Parametrizations of the meson-nucleon couplings}

The parameters of the model include  nucleon, ($M=939 MeV$), and 
meson
($m_{\sigma}$, $m_{\omega}$, $m_{\rho}$, $m_{\delta}$, see Table 1) masses
 and the density dependent
 meson-nucleon couplings.
The density dependence parametrization used here, inspired by 
$DB$ calculations 
\cite{jong98,hof01,vandal04},
was proposed \cite{TW99,gait04,avan04} as :

\begin{equation}\label{eq:30}
g_{i}(\rho)=g_{i}(\rho_0)f_{i}(x), ~~~~for ~~i=\sigma,\omega,\rho,\delta,
\end{equation}

\noindent
with

\begin{eqnarray}\label{eq:31}
&&f_{i}(x)=a_{i}\frac{1+b_{i}(x+d_i)^2}{1+c_{i}(x+d_i)^2},~
i=\sigma,\omega, \nonumber \\
&&f_{i}(x)=a_{i}exp[-b_{i}(x-1)] - c_{i}(x-d_i),i=\rho,\delta
\end{eqnarray}

\noindent
where $x=\rho/\rho_0$ and $\rho_0$ is the saturation density.

\noindent
Parametrization form and parameters are taken from 
ref.\cite{TW99} 
for $\sigma$, $\omega$ mesons and from ref.\cite{gait04,avan04} 
for $\rho$, $\delta$ 
mesons, respectively.
All parameters are listed in Table 1.
 The density dependent couplings as a function of baryon density are 
displayed in Fig. \ref{gcoup}.

\begin{figure}[hbtp]
\begin{center}
\includegraphics[scale=0.43]{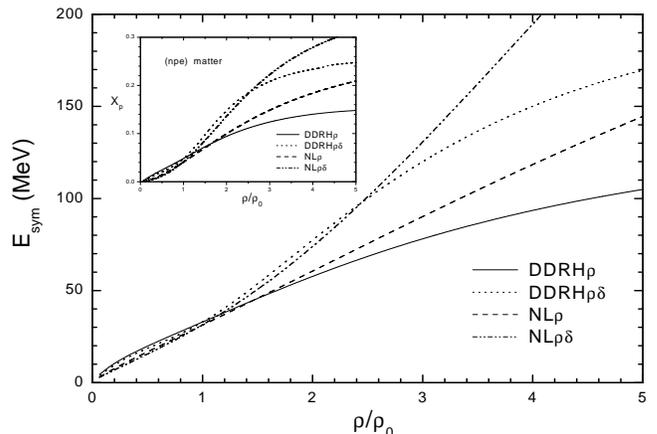}
\vglue -0.5cm
\caption{Density dependence of symmetry energy. Insert: proton fraction of
the corresponding $\beta$-equilibrium ($npe$) system.}
\label{ensym}
\end{center}
\end{figure}

For symmetric matter at saturation ($\rho_0=0.153~fm^{-3}$) we get a binding
energy $E/A=\epsilon/\rho-M=-16.25~MeV$ and a compressibility modulus
$K=240 MeV$. In order to remark the effects of the coupling density dependence
we will compare the results with a non-linear ($NL$) relativistic mean field 
model {\it with constant couplings} which presents very similar saturation 
properties (Set A of ref.\cite{liubo05}), including a symmetry 
energy $E_{sym}=31.3~MeV$.
Both effective models, $NL$ and $DDRH$ are rather soft for symmetric matter 
at high density, in agreement with relativistic collision data, 
\cite{dan00,dan02}, and
Dirac-Brueckner expectations \cite{gross99,gait01}. 

As shown in refs.\cite{liubo02,liubo05} when we 
include the $\delta$ coupling we have to increase the $\rho$ coupling in 
order to keep the same symmetry term at saturation (see Table 1).
Since at higher densities the $\delta$ coupling is increasing while the
$\rho$ one is decreasing (see Fig.\ref{gcoup}), as a result in the $DDRH$
choice the 
symmetry term will be
less repulsive than in the $NL$ case. This can be clearly seen in 
Fig. \ref{ensym} at densities above $2.5 \rho_0$.
In the insert we present the corresponding proton fraction in a 
$\beta$-equilibrated $npe$ system.

\begin{figure}[hbtp]
\begin{center}
\includegraphics[scale=0.45]{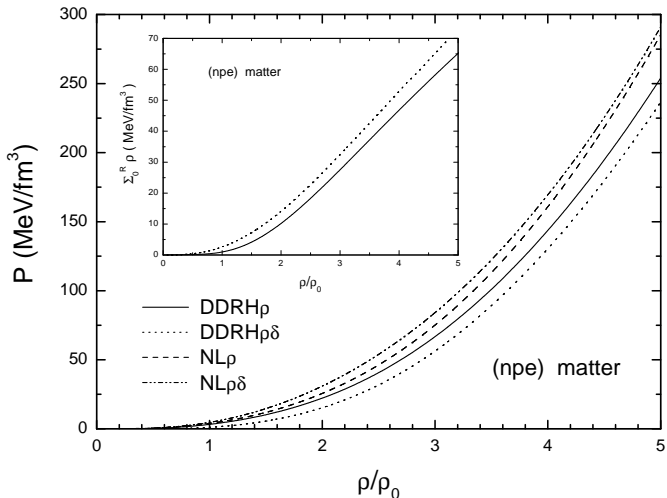}
\vglue -0.5cm
\caption{Equation of state for ($npe$) matter in different models.
 Insert: density dependence of the rearrangement terms in the $DDRH$ cases}
\label{pnpe}
\end{center}
\end{figure}

\begin{figure}[hbtp]
\begin{center}
\includegraphics[scale=0.38]{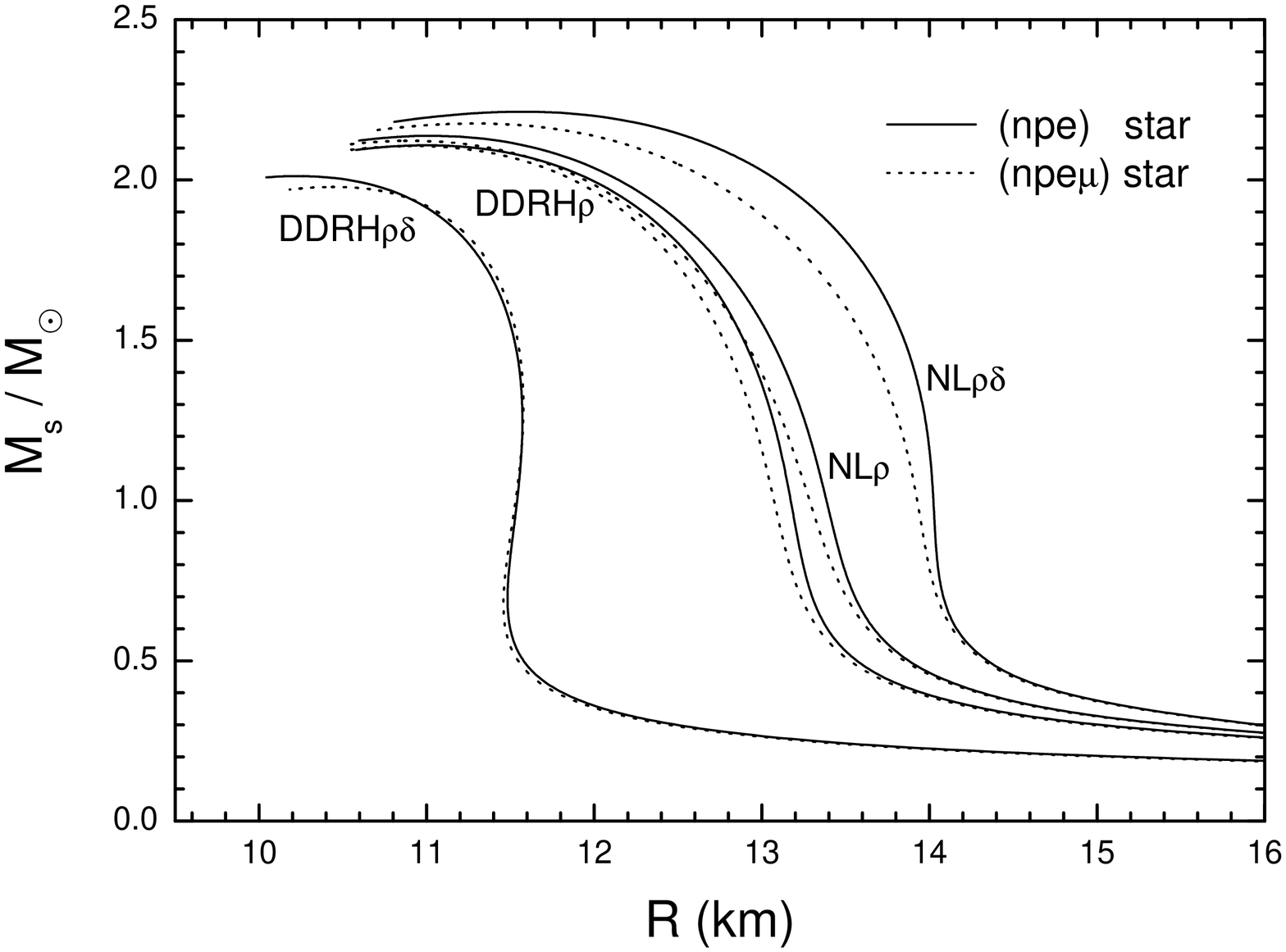}
\vglue -0.5cm
\caption{Mass of the neutron star as a function of the radius of the 
neutron star in the two models.}
\label{smr}
\end{center}
\end{figure}

\vspace{-1.0cm}
\section*{4. Neutron star results}

The $\beta$-equilibrium nuclear matter is relevant for the composition of the 
neutron stars, as discussed in the previous Section..
The $EOS$, pressure vs. density, for ($npe$) matter in the density 
dependent $DDRH$ vs.$NL-RMF$ models is reported in Fig. \ref{pnpe}.
 We see that, at variance 
with the $NL$ results, in the 
$DDRH$ cases the $EOS$ without $\delta$-meson is stiffer than
 that with the $\delta$-meson. This is partially due to the softening of 
the symmetry term in the $DDRH\rho\delta$ choice joined to a larger negative 
contribution to the pressure from the rearrangement term, see Eq.(\ref{eq:14}),
 as shown inside Fig. \ref{pnpe}. We note that both effects are related
to the density increase of the effective $g_\delta$ coupling 
(see Fig.\ref{gcoup}) as expected from Dirac-Brueckner calculations
\cite{jong98,hof01,vandal04}.

We use the two effective nucleon-meson lagrangians, with and without 
density dependent couplings, to calculate neutron star ($NS$) properties, with
particular attention to the $\delta$-field effects.
The correlation between neutron star mass and radius for
the $\beta$-equilibrium ($npe$) and $(npe\mu)$ matter obtained by the  
$DDRH$ (density dependent) and $NL-RMF$(constant couplings) 
parametrizations  are shown in Fig. \ref{smr}.
The obtained maximum mass, corresponding radius and central density for
 the ($npe$) and $(npe\mu)$ neutron star matter are reported in Table 2.
 
We first note that the $NL\rho$ and $DDRH\rho$ results are rather similar,
with the $DDRH\rho$ interaction leading to a little softer matter, slightly 
smaller $NS$ mass $M_S$ and radius $R$ and larger central 
density (see Table 2).
When we include the $\delta$ coupling we observe a clear effect in opposite 
directions: the $DDRH$ case becomes much softer while the $NL-RMF$ 
choice shows a much stiffer behavior. This can be seen from Table 2, for the 
variations in $M_S/R$ and central densities, but in fact it is quite impressive
as it appears in Fig.\ref{smr}: with reference to the close $DDRH\rho/NL\rho$
curves we see a clear shift to the ``left'' of the $DDRH\rho\delta$
predictions and just the opposite to the ``right'' for the $NL\rho\delta$
expectations. 

\begin{table*}
\begin{center}
{{\large \bf Table 2.}~~Maximum mass, corresponding radius and
central density of the star by the different models}.
\par \noindent
\vspace{0.3cm}
\begin{tabular}{c|c|c|c|c|c} \hline
$        $      &$Model$     &\multicolumn{2}{|c|}{Dens.Dip} 
 &\multicolumn{2}{c}{$RMF$}\\ \hline
$neutron~star$  &$properties$        &$DDRH\rho$ &$DDRH\rho\delta$ 
&$NL\rho$ &$NL\rho\delta$ \\ \hline
$(npe)~matter$  &$M_{S}/M_{\bigodot}$ &2.108     &2.01            
&2.14   &2.21 \\ \cline {2-6}
                &$R (km)$             &11.00     &10.29           
&11.02  &11.55\\ \cline {2-6}
                &$\rho_c/\rho_0$      &6.99      &7.41            
&6.78   &6.44\\ \hline

$(npe\mu)~matter$ &$M_{S}/M_{\bigodot}$ &2.106   &1.98            
&2.12   &2.18\\ \cline {2-6}
                  &$R (km)$             &10.91   &10.27           
&10.91  &11.30\\ \cline {2-6}
                  &$\rho_c/\rho_0$      &7.14    &7.44            
&6.93   &6.71\\ \hline
\end{tabular}
\end{center}
\vspace{-0.3cm}
\end{table*}

In general we also see, in particular from Table 2, that the ($npe$) star 
matter, for all models, has slightly larger masses and radii, and lower 
central densities, than the $(npe\mu)$ star matter. This is due to the fact 
that the $(npe\mu)$ star matter has some larger proton fraction in the
regions above a critical baryon density where the muon appears, 
as already noted at the end of Sect.2.

\vspace{-0.5cm}
\section*{5. Conclusion and outlook}
All microscopic approaches of Dirac-Brueckner type to an effective 
meson-nucleon Lagrangian picture of the nuclear matter are predicting a 
density dependence of the couplings. We have studied the relative effects 
on the nuclear 
$EOS$ at high baryon and isospin density, with application to nucleon-lepton
neutron star properties. In particular we have focussed our attention on the 
contribution of the isovector-scalar $\delta$-meson. In fact in the 
``constant coupling'' ($NL-RMF$) scheme the $\delta$ leads to very repulsive 
symmetry energy at high density. At variance in the ``density dependence'' 
case ($DDRH$) we can have a ``softer'' dense asymmetric matter due to 
combined mechanism of a decrease
of the isovector-vector $g_\rho$ coupling and an increase of the 
$g_\delta$ (isovector scalar), which even leads to a larger pressure 
reduction from the
 rearrangement terms.
The effect is clearly seen on equilibration properties of ($npe$) and/or 
($npe\mu$) neutron stars, with an interesting decrease of the $NS$ mass in 
the $DDRH$ case when the $\delta$ contribution is included. We note that  
pure nucleon-lepton models cannot easily predict $NS$ masses below two
solar units. Our results seem to indicate that the large uncertainty of 
nucleon matter predictions, see
the recent review \cite{page06}, of relevance even for hybrid
quark models, can be associated to the density dependence of the effective 
meson-nucleon couplings, in particular of the  $g_\delta$.

In conclusion we remark the interest of future work on two main 
directions:

i) The importance of further $DB$ confirmations of the high density behavior of
the meson-nucleon effective couplings, in particular of some fundamental ground
for the expected increase of the  $g_\delta$;

ii) The study of dynamical effects of the isovector meson fields at the high 
baryon and isospin densities that can be reached in relativistic heavy ion 
collisions with exotic beams. Differential flows and particle productions 
appear to be rather promising observables, see the recent refs.
\cite{bao05,qli0506,qli06,fer06,dit07}. 
\vspace{-1.0cm}
\section*{Acknowledgments}

We would like to thank H.-J. Schulze for helpful discussions.
This project is supported by the National Natural Science Foundation of China
under Grant No.10275002,  Grant No.10575005 and Grant No.10675046,
the Natural Science Foundation of Zhejiang Province of China under Grant 
No.Y605476, and the INFN of Italy.


\end{document}